\documentclass{ws-p9-75x6-50}

\newcommand{\hmpc}{$h^{-1}$Mpc}

\begin{document}

\title{Quantitative Analysis of the Void Size Distribution in the LCRS 
and in CDM Models}

\author{V. M\"uller, S. Arbabi-Bidgoli}

\address{Astrophysikalisches Institut Potsdam, D-14482 Potsdam, Germany\\
E-mail: vmueller@aip.de, sarbabi@aip.de}


\maketitle

\abstracts{
We have analyzed the size distribution of voids in 2D slices of the Las Campanas
Redshift Survey (LCRS).  The characteristic sizes show a scaling $D = D_0 +
\nu\lambda$ with the mean galaxy separation $\lambda$.  Comparison with mock
samples of the LCRS in 2D and 3D, using various simulations, cosmologies and
galaxy identification schemes, gives a similar scaling, but with steeper slope
and a lack of large voids.  Best results are obtained for dark matter halos in a
$\Lambda$CDM model.}

\section{Voids in the LCRS}

The appearence of large voids in the galaxy distribution of extended galaxy
surveys as the Las Campanas Redshift Survey (LCRS, Shectman et
al.~\cite{LCRS96}) was one of the most spectacular findings characterizing the
large-scale matter distribution of the universe.  We expect large voids due to
the matter outflow from positive primordial potential
perturbations~\cite{MDGM98,LS98}.  A second mechanism for the lack of galaxies
in voids is the lower probability of galaxy formation in underdense regions
known as biasing.  Recently, Friedmann and Piran~\cite{FP00} presented a simple
model for void formation based on the probability distribution of spherical
underdense regions in the universe, and the less abundance of density peaks
therein.  There it was aimed to explain that in recent redshift surveys voids
have radii between 13 and 30 \hmpc, occupy 50\% of the volume, and that the
probability of larger voids falls off exponentially.  Since the theoretical
conclusions are based on a simple analytical model, and since galaxy surveys
before LCRS were not extended enough to get reliable void statistics, we have
undertaken a new quantitative study of voids in the LCRS and in numerical
simulations~\cite{MAET00}.  Here this analysis is evaluated and compared with
theoretical expectations.

In our analysis~\cite{MAET00} we have used automatic void finders in 2D.  Voids
are identified as connected empty cells on a grid, where first square base voids
are identified, and then boundary layers are added if they are empty and larger
than 2/3 of the previous extension.  The void finder starts with the search for
the largest voids.  The algorithm was first described and tested in 2D using
slices through the CfA galaxy redshift survey by Kauffmann \&
Fairall~\cite{KF91}.

\begin{figure}[t] 
\epsfxsize=35pc 
\vspace{-1cm}
\hspace{-2cm}
\epsfbox{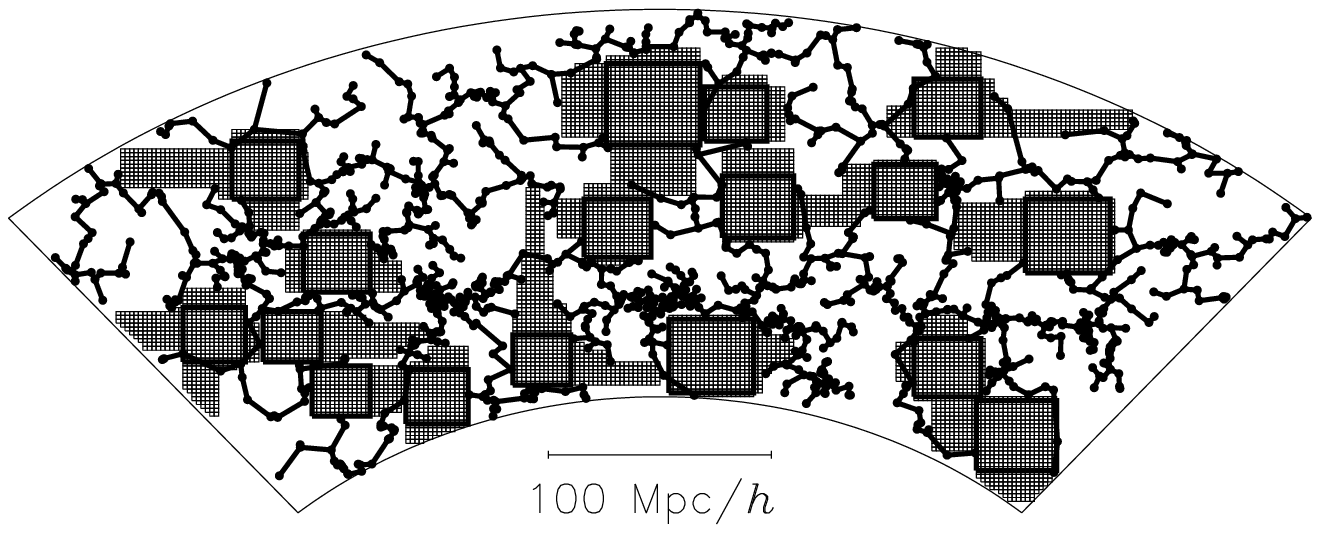} 
\vspace{-2cm}
\caption{Large voids in a well sampled LCRS slice.  Voids are empty regions
marked by square base voids and rectangular extensions.  Galaxies are shown by
dots connected with the minimal spanning tree to emphasize the overdense
regions}
\label{f1}
\end{figure}

From the LCRS we have selected 14 volume limited subsets in different radial
ranges of the six narrow slices, projected them into the central planes and
corrected for variations in the sampling fraction.  In Fig.~\ref{f1} the void
distribution of such sets is shown which illustrates that we got a large number
of voids in each data set, and that the voids cover smoothly the surface of the
slice, with no obvious inhomogeneities or biasing by boundary effects or radial
gradients.  To emphasize the importance of large voids, we constructed the size
distribution as cumulative coverage of the survey area with voids.  Results are
shown in Fig.~\ref{f2}.  In the left pannel we overplot the void distributions
of all 14 samples in normalizing them with the median void sizes from each
sample.  The thick lines show the results from the best sampled part of the
survey.  Altogether we get a nice convergence to a mean behavior shown by a
smooth parametric fit.  Also the derivative is shown in comparison with the
distribution of voids in random point distributions which shows the main result
of the gravitational clustering of galaxies in overdense regions:  The relative
abundence both of large voids and of small voids in the data is larger than in
the random point distributions (the dashed lines).

\section{Comparison with simulations}

We have compared the LCRS voids with cold dark matter simulations in spatially
flat $\Lambda$-term cosmology ($\Lambda$CDM, $\Omega_0=0.3$) and in open models
(OCDM, $\Omega_0=0.5$).  PM simulations were performed in 500 $h^{-1}$Mpc boxes
using $300^3$ particles and a threshold bias prescription to identify `galaxies'
with single particles.  In addition we analyzed P3M simulations in 280
$h^{-1}$Mpc boxes and with $256^3$ particles, a friend-of-friend halo finder, and a
procedure to split off unvirialized halos that result from particle clumps
merged by numerical effects.  The mock samples in the different simulations fit
well the 2-point correlation function of the LCRS as already shown in Tucker et
al.~\cite{Tu97}.  In the right pannel of Fig.~\ref{f3}, we compare the real
space correlation function of the mock samples with the reconstructed 3D
correlation function of APM galaxies~\cite{B96}.  The mock samples provide good
fits to the observations especially on large scales.

\begin{figure}[t] 
\epsfxsize=30pc 
\epsfbox{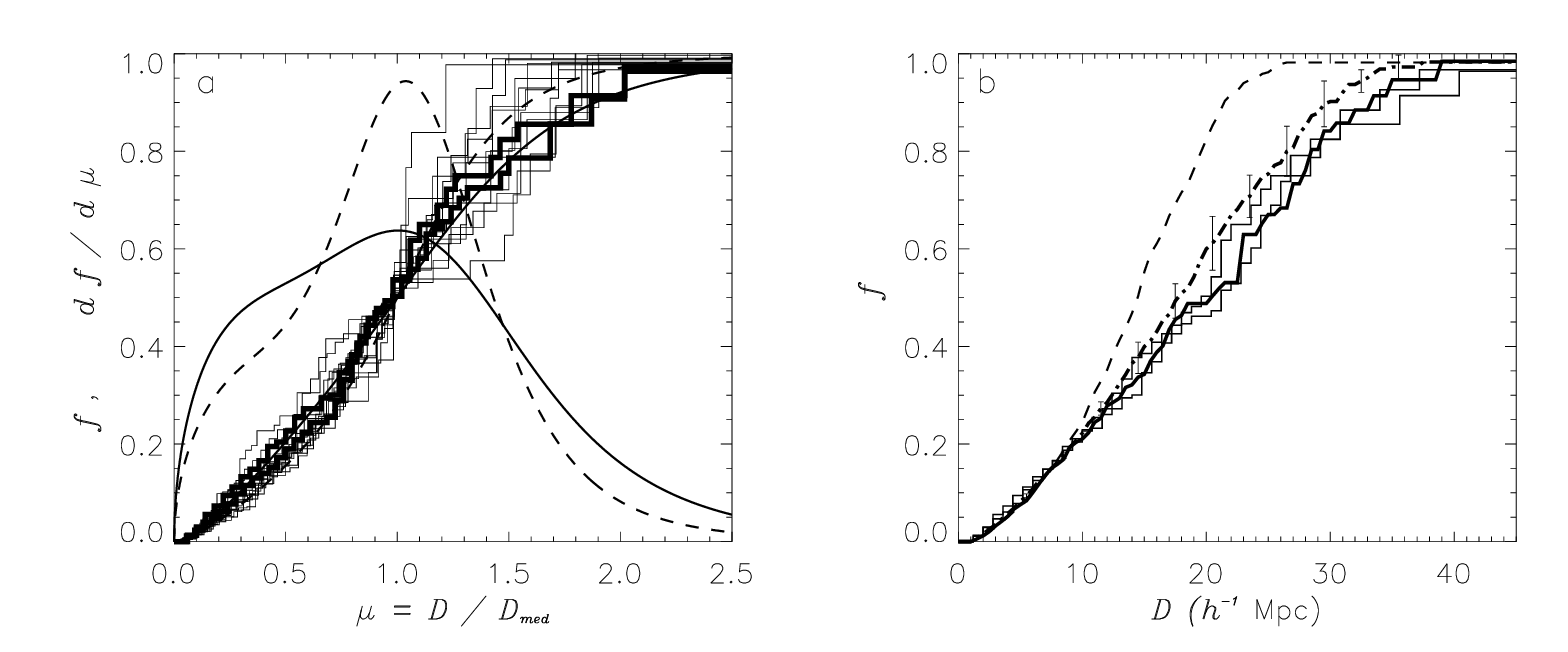} 
\caption{Cumulative area covered by voids as function of the size $D$ compared
to voids in a random point distribution ({\it dashed lines}).  ({\bf a}) LCRS
void sizes measured in terms of the median size together with a fit ({\it solid
line\,}) and its derivative.  ({\bf b}) LCRS histograms compared to mock samples
of OCDM ({\it dash-dotted line with errors}) and $\Lambda$CDM ({\it thick line})}
\label{f2} 
\end{figure}

The cumulative void size distribution of ten OCDM mock samples is shown in the
right pannel of Fig.~\ref{f2} together with the variance range.  Obviously this
model has difficulty in matching the void distribution for a large range from 15
to 35 \hmpc.  The $\Lambda$CDM fits better, but it also seems to underestimate
the size of the largest voids.

\section{Scaling Relation}

The 14 analyzed volume-limited data sets show variations in the galaxy densities
which are due to different absolute magnitude cuts and different sampling
fractions.  We already showed~\cite{MAET00} that differences in the magnitude
ranges have negligible influence on the size distribution of voids.  So we
ascribed the differences in the parameters of the void distributions to
variations in the galaxy sampling densities.  We measured it by the mean galaxy
separation $\lambda$.  In analyzing both data and mock samples, we found a
characteristic scaling relation for the mean void sizes that cover 25\%, 50\%
and 75\% of the survey area.  This scaling is a simple linear relation $D = D_0
+ \nu\lambda$, with size $D_0=(6,12,17)\,h^{-1}{\mbox{Mpc}}$ and slope
$\nu=(0.9,1.1,1.5)$ for 25\%, 50\% and 75\% of void coverage, cp.
Table~\ref{voids}.

\begin{figure}[t] 
\epsfxsize=30pc 
\epsfbox{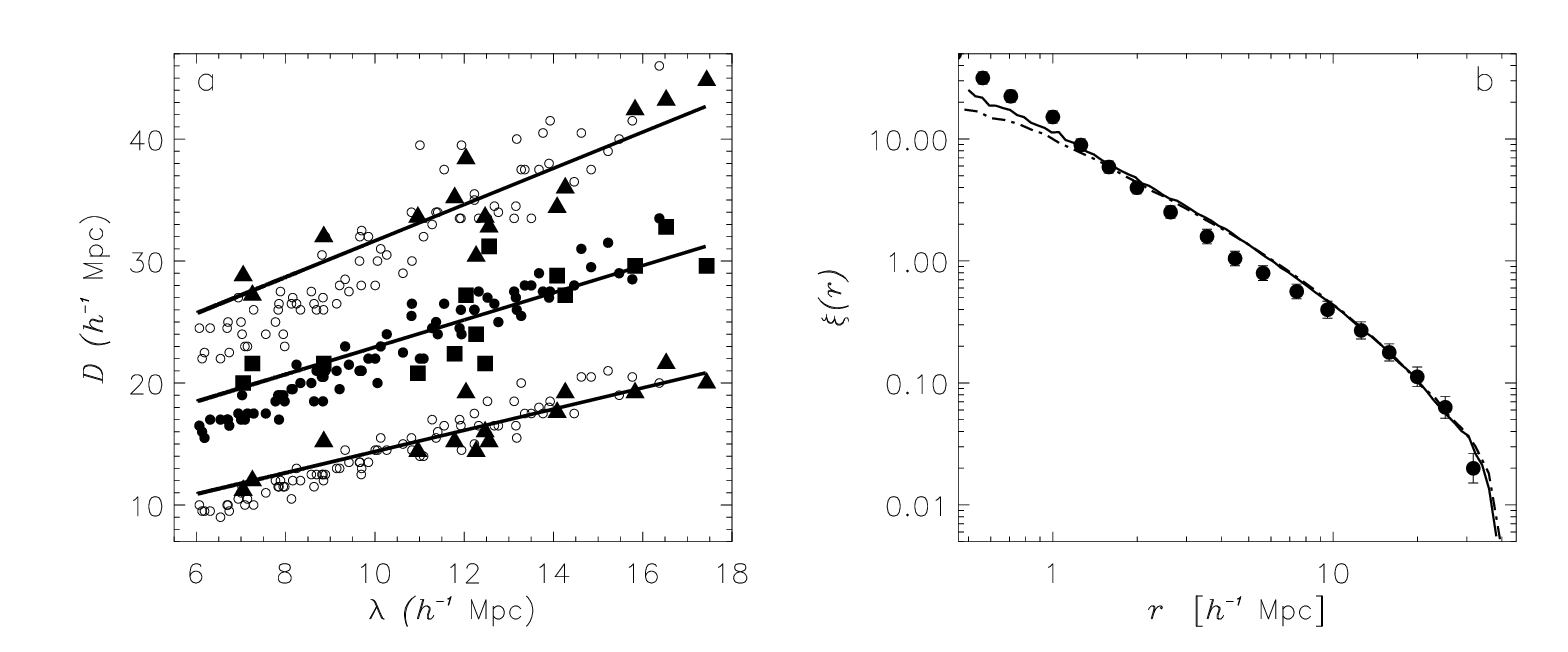} 
\caption{({\bf a}) Median ({\it squares}) and quartiles ({\it triangles}) of 
the void sizes versus the galaxy separation $\lambda$ and the scaling relations 
as thick lines. The circles ({\it filled for the median}) for the OCDM 
mock samples show smaller void sizes for low $\lambda$. 
({\bf b}) Comparison of the real space two-point correlation function in the APM 
survey compared with the $\Lambda$CDM (solid line) and OCDM (dashed-dotted line) 
mock samples.}
\label{f3} 
\end{figure}

In Fig.~\ref{f3} we show that the scaling relations in simulated OCDM mock
samples are steeper, $\nu_{model} > \nu_{data}$.  It is even steeper for the
$\Lambda$CDM model as seen in Table~\ref{voids}.  According to our analysis, the
median size of voids is larger than reported from earlier surveys~\cite{FP00}.
The largest voids have a size of $D_{max}\approx 50\,h^{-1}{\mbox{Mpc}}$ not met
in simulations.  Tests with mock samples indicate that the 2D-void size
distributions in LCRS slices have a fixed ratio to the 3D-voids.  Further, we
have shown that large voids increase in size by 10 \% in redshift space, but
small voids remain unchanged, cp.  also~\cite{SRM01}.  Our mock samples were
always analyzed in redshift space.

\begin{table}[t]
\caption{Parameters of the similarity relation of the void size distribution. 
\label{voids}}
\begin{center}
\footnotesize
\begin{tabular}{|c|cc|cc|rc|}
\hline
{\rule[-1pt]{0mm}{3ex} data} & 
 \multicolumn{2}{|c|}{median} & \multicolumn{2}{|c|}{lower quartile} & 
 \multicolumn{2}{|c|}{upper quartile} \\
\hline
{\rule[-1pt]{0mm}{3ex}} 
     & $D_0$ & $\nu$ & $D_0$ & $\nu$ & $D_0$ & $\nu$ \\
     & \hmpc &       & \hmpc &       & \hmpc &       \\
\hline
{\rule[-1pt]{0mm}{3ex}} 
LCRS     & $11.8 \pm 2.9$ & $1.1 \pm 0.2$ & $5.7 \pm 1.6$ & $0.9 \pm 0.1$ 
         & $16.8 \pm 2.9$ & $1.5 \pm 0.2$ \\
                  
Poisson  &  $0.9 \pm 1.0$ & $1.8 \pm 0.1$ & $0.9 \pm 0.3$ & $1.2 \pm 0.1$ 
         &  $2.1 \pm 0.5$ & $2.1 \pm 0.1$ \\ 
$\Lambda$CDM &  $7.6 \pm 0.9$ & $1.5 \pm 0.1$ & $2.9 \pm 1.2$ & $1.1 \pm 0.1$ 
         & $14.2 \pm 1.2$ & $1.8 \pm 0.1$ \\
OCDM     & $10.7 \pm 1.8$ & $1.2 \pm 0.2$ & $4.8 \pm 0.6$ & $1.0 \pm 0.1$ 
         & $15.7 \pm 2.7$ & $1.8 \pm 0.2$ \\
\hline
\end{tabular}
\end{center}
\end{table}

\section{Discussion}

The void statistics provide important characteristics of the galaxy distribution
in low to medium dense regions where the correlation function is less sensitive.
The self-similarity of the size distribution of samples with different mean
galaxy separations culled by random dilutions and the scaling relation of the
percentile sizes represent a genuine feature of hierarchical clustering.  The
discrepancy of the simulated void distribution to observations requires a
stronger biasing of simulated galaxies and a higher concentration of them in
superclusters, as has also been seen by analyzing overdensity regions with the
minimal spanning tree technique illustrated in Fig.~\ref{f1}, cp.  Doroshkevich
et al.~\cite{DMRT99}.


\begin{thebibliography}{99}

\bibitem{LCRS96} S.~A. Shectman, S.~D. Landy, A. Oemler, D.~L. Tucker, H. Lin, 
R.~P. Kirshner, P.~L. Schechter , {\sl ApJ} {\bf 470}, 172 (1996).

\bibitem{MDGM98} S. Madsen, A.G. Doroshkevich, S. Gottl\"ober, V. M\"uller, 
{\sl A\&A} {\bf 329}, 1 (1998).

\bibitem{LS98} J. Lee, S. Shandarin, {\sl ApJ} {\bf 505}, L75 (1998).

\bibitem{FP00} Y. Friedmann, T. Piran, {\sl ApJ} in press, astro-ph/0009320 
(2000).

\bibitem{MAET00} V. M\"uller, S. Arbabi-Bidgoli, J. Einasto, D. Tucker, 
{\sl MNRAS} {\bf 325}, 280 (2000).

\bibitem{SRM01} 
J.D. Schmidt, B.S. Ryden, A.L. Melott, {\sl ApJ} {\bf 546}, 609 (2001).

\bibitem{KF91} G. Kauffmann, A.P. Fairall 1991, {\sl MNRAS} {\bf 248}, 313 (1991).

\bibitem{Tu97} D.T. Tucker, A. Oemler, R.P. Kirshner, H. Lin, S.A. Shectman,
S.D. Landy, P.L. Schechter, V. M\"uller, S. Gottl\"ober, J. Einasto,  
{\sl MNRAS} {\bf 285}, L5 (1997).

\bibitem{B96} C.M. Baugh, {\sl MNRAS} {\bf 280}, 267 (1996).

\bibitem{DMRT99} A.G. Doroshkevich,  V. M\"uller, J. Retzlaff, V. Turchaninov, 
{\sl MNRAS} {\bf 306}, 575 (1999).

\end{thebibliography}
\end{document}